\documentclass{article}
\usepackage{spconf,amsmath,graphicx,subcaption,amssymb,multirow,stfloats}
\usepackage[hang,flushmargin]{footmisc}

\newcommand\blfootnote[1]{%
  \begingroup
  \renewcommand\thefootnote{}\footnote{#1}%
  \addtocounter{footnote}{-1}%
  \endgroup
}

\title{TRANSFORMER-BASED VARIABLE-RATE IMAGE COMPRESSION WITH REGION-OF-INTEREST CONTROL}
\name{Chia-Hao Kao$^{\star}$ \quad Ying-Chieh Weng$^{\star}$\thanks{$^{\star}$ Equal contribution.} \quad Yi-Hsin Chen \quad Wei-Chen Chiu \quad Wen-Hsiao Peng}
\address{National Yang Ming Chiao Tung University, Taiwan}
\usepackage{overpic}
\usepackage{enumitem} 
\usepackage{overpic} 
\usepackage{color}

\definecolor{turquoise}{cmyk}{0.65,0,0.1,0.3}
\definecolor{purple}{rgb}{0.65,0,0.65}
\definecolor{dark_green}{rgb}{0, 0.5, 0}
\definecolor{orange}{rgb}{0.8, 0.6, 0.2}
\definecolor{red}{rgb}{0.8, 0.2, 0.2}
\definecolor{darkred}{rgb}{0.6, 0.1, 0.05}
\definecolor{blueish}{rgb}{0.0, 0.3, .6}
\definecolor{light_gray}{rgb}{0.7, 0.7, .7}
\definecolor{pink}{rgb}{1, 0, 1}
\definecolor{greyblue}{rgb}{0.25, 0.25, 1}

\definecolor{light-gray}{gray}{0.70}

\begin{document}
%
\maketitle
\begin{abstract}
This paper proposes a transformer-based learned image compression system. It is capable of achieving variable-rate compression with a single model while supporting the region-of-interest (ROI) functionality. Inspired by prompt tuning, we introduce prompt generation networks to condition the transformer-based autoencoder of compression. Our prompt generation networks generate content-adaptive tokens according to the input image, an ROI mask, and a rate parameter. The separation of the ROI mask and the rate parameter allows an intuitive way to achieve variable-rate and ROI coding simultaneously. Extensive experiments validate the effectiveness of our proposed method and confirm its superiority over the other competing methods.  
\blfootnote{This work was supported by National Science and Technology Council, Taiwan under Grants NSTC 111-2634-F-A49-010 and MOST 110-2221-E-A49-065-MY3, MediaTek Advanced Research Center, and National Center for High-performance Computing.}

\end{abstract}

\section{Introduction}
\label{sec:intro}
Transformers have recently emerged as an attractive alternative to convolutional neural networks (CNN) for constructing learned image compression systems~\cite{lu2022transformer, zhu2022transformerbased}. The attention-based convolution coupled with the shifted-windowing technique~\cite{liu2021swin} offers both high compression performance and low computational cost. To make learned image codecs practical, much research has been devoted to the use of a single autoencoder for variable-rate compression. However, little work is done on transformer-based codecs. 

One common approach to variable-rate compression with a single autoencoder is to adapt the feature distributions of the autoencoder. For example, Yang et al.~\cite{yang2020variable} channel-wisely scale the feature maps of every convolutional layer in the autoencoder according to a rate parameter. In contrast, Cui et al.~\cite{cui2020gained} perform channel-wise scaling for the image latents only. Specifically, they first optimize the scaling factors for a few distinctive rate points and then interpolate between the resulting scaling factors for continuous rate adaptation. 
In a similar vein, Wang et al.~\cite{wang2022block} scale the value matrix channel-wisely in each self-attention layer for transformer-based codecs. 
In another direction, Song et al.~\cite{song2021variable} propose a spatially adaptive rate adaptation scheme by introducing spatial feature transform (SFT)~\cite{wang2018recovering} to convolutional layers for an element-wise affine transformation of the feature maps. In particular, the affine parameters are predicted by a conditioning network that takes as input a quality map reflecting the spatial importance of every image pixel. Notably, this quality map can be adapted for multiple uses, such as rate control, spatial bit allocation, region-of-interest (ROI) coding, and task-specific coding. However, how to determine the quality map is non-trivial and may involve time-consuming back-propagation at inference, especially when it is necessary to combine some of these tasks, e.g. ROI coding subject to a rate constraint.

In this work, we propose a transformer-based image codec capable of achieving variable-rate compression while supporting ROI functionality. Inspired by prompting techniques~\cite{liu2021prompt, jia2022visual}, we introduce prompt generation networks to condition our transformer-based codec. Unlike ordinary prompting~\cite{liu2021prompt, jia2022visual}, which learns task-specific prompt tokens, our prompt generation networks generate content-adaptive tokens according to the input image, an ROI mask, and a rate parameter. The separation of the ROI mask and the rate parameter allows us to disentangle the rate and spatial quality controls. Our contributions are threefold. (1) To our best knowledge, this work is the first transformer-based image codec that leverages network-generated prompts to achieve variable-rate coding with ROI support. (2) Our scheme offers an intuitive way to specify the ROI and rate parameters. (3) Our scheme performs comparably to or better than the baselines while having lower computational complexity.

\begin{figure*}[t]
    \centering
    \includegraphics[width=0.88\textwidth]{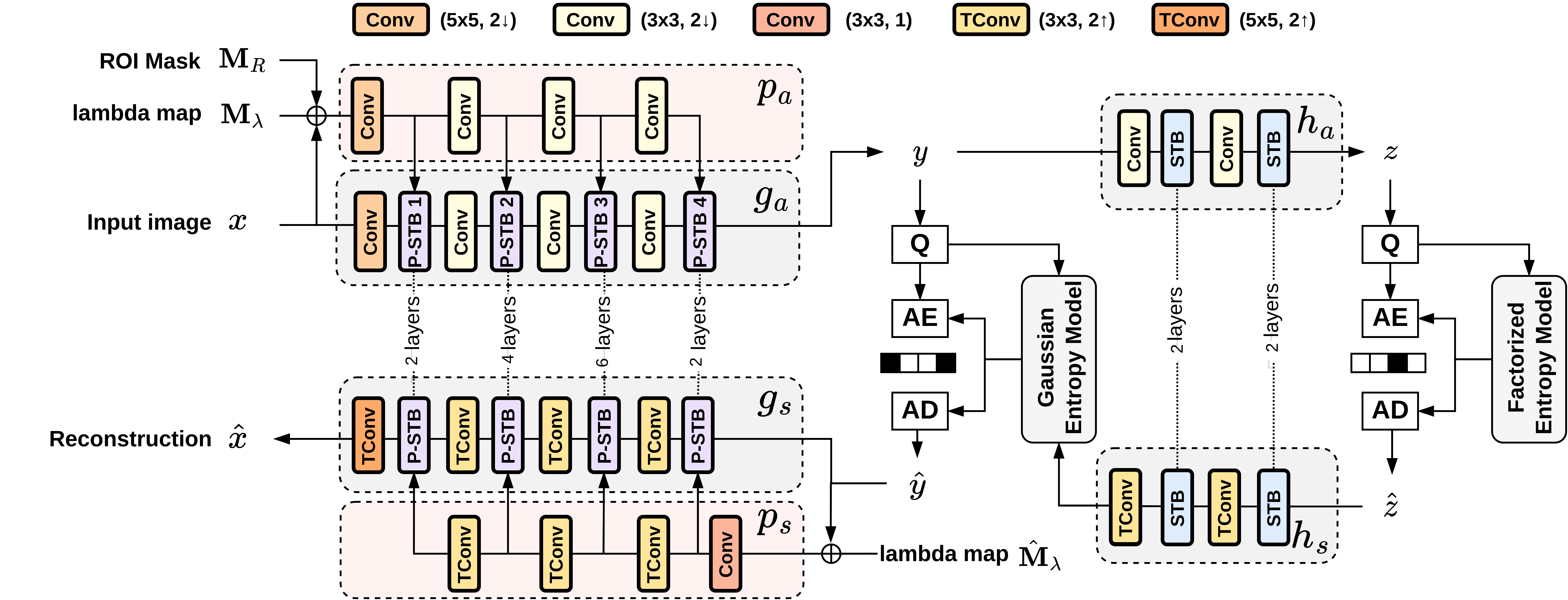}
    \caption{The network architecture of the proposed transformer-based image codec. For simultaneous rate and spatial quality control, the prompt generation networks $p_a, p_s$ produce prompt tokens for the encoder $g_a$ and decoder $g_s$, respectively. 
    }
    
    \label{fig:architecture}
    \vspace{-0.3cm}
\end{figure*}

\section{Proposed Method}
\label{sec:method}

We propose a Swin-transformer-based image compression system. It is capable of performing variable-rate compression with a single model while offering spatially adaptive quality control for the ROI functionality. Fig.~\ref{fig:architecture} illustrates our overall architecture. It is built upon TIC (\textbf{T}ransformer-based \textbf{I}mage \textbf{C}ompression~\cite{lu2022transformer}) but without the context model for entropy coding. The main autoencoder $g_a, g_s$ and hyperprior autoencoder $h_a, h_s$ are comprised of Swin-transformer blocks (STB) interleaved with convolutional layers. The details of STB can be found in~\cite{lu2022transformer}. To encode an input image $x\in\mathbb{R}^{3\times H\times W}$, the encoder takes two additional inputs, a lambda map $\mathbf{M}_\lambda\in\mathbb{R}^{1\times H \times W}$ and a ROI mask $\mathbf{M}_R\in\mathbb{R}^{1\times H \times W}$. The lambda map $\mathbf{M}_\lambda$ is a uniform map populated with the same rate parameter $m_\lambda\in[0,1]$ that controls the bit rate of the compressed bitstream. The ROI mask $\mathbf{M}_R$ specifies spatially the importance of individual pixels in the image. Each element in the ROI mask is a real value in $[0,1]$. Both inputs serve as the conditioning signal utilized to generate \textit{prompt tokens} for adapting the main encoder $g_a$ (Section~\ref{sec:prompt}). In a similar way, the decoder $g_s$ is adapted by taking as inputs the quantized image latent $\hat{y}$ and a downscaled lambda map $\hat{\mathbf{M}}_\lambda \in \mathbb{R}^{1\times \frac{H}{16}\times \frac{W}{16}}$ that matches the spatial resolution of the latent $\hat{y}$. Unlike~\cite{song2021variable}, which relies on a (single) quality map for both rate and spatial quality control, our design has the striking feature of disentangling the rate (i.e.~$\mathbf{M}_\lambda$) and spatial quality control (i.e.~$\mathbf{M}_R$). In other words, it treats them as two independent dimensions, offering a more intuitive way to achieve simultaneous variable-rate and ROI coding. 


\subsection{Prompt-based Conditioning}
\label{sec:prompt}
Inspired by~\cite{jia2022visual}, we propose to use learned parameters, known as prompts, as the additional inputs to the Swin-transformer layers, in order to achieve variable-rate and ROI coding. The resulting STB is termed prompted Swin-transformer block (P-STB). As shown in Fig.~\ref{fig:architecture}, the learned prompts are produced by two generation networks $p_a, p_s$ for conditioning the encoder $g_a$ and decoder $g_s$, respectively. $p_a$ consists of several convolutional layers that match those of the encoder $g_a$, and it takes as input the concatenation of the ROI mask $\mathbf{M}_R$, lambda map $\mathbf{M}_\lambda$, and image $x$. The feature maps of $p_a$ are fed into the corresponding P-STBs to generate prompt tokens to be interacted with image tokens. $p_s$ follows a similar architecture, replacing the convolutional layers with the transposed convolutional layers for upsampling. 

\begin{figure}
    \centering
    \includegraphics[width=0.46\textwidth]{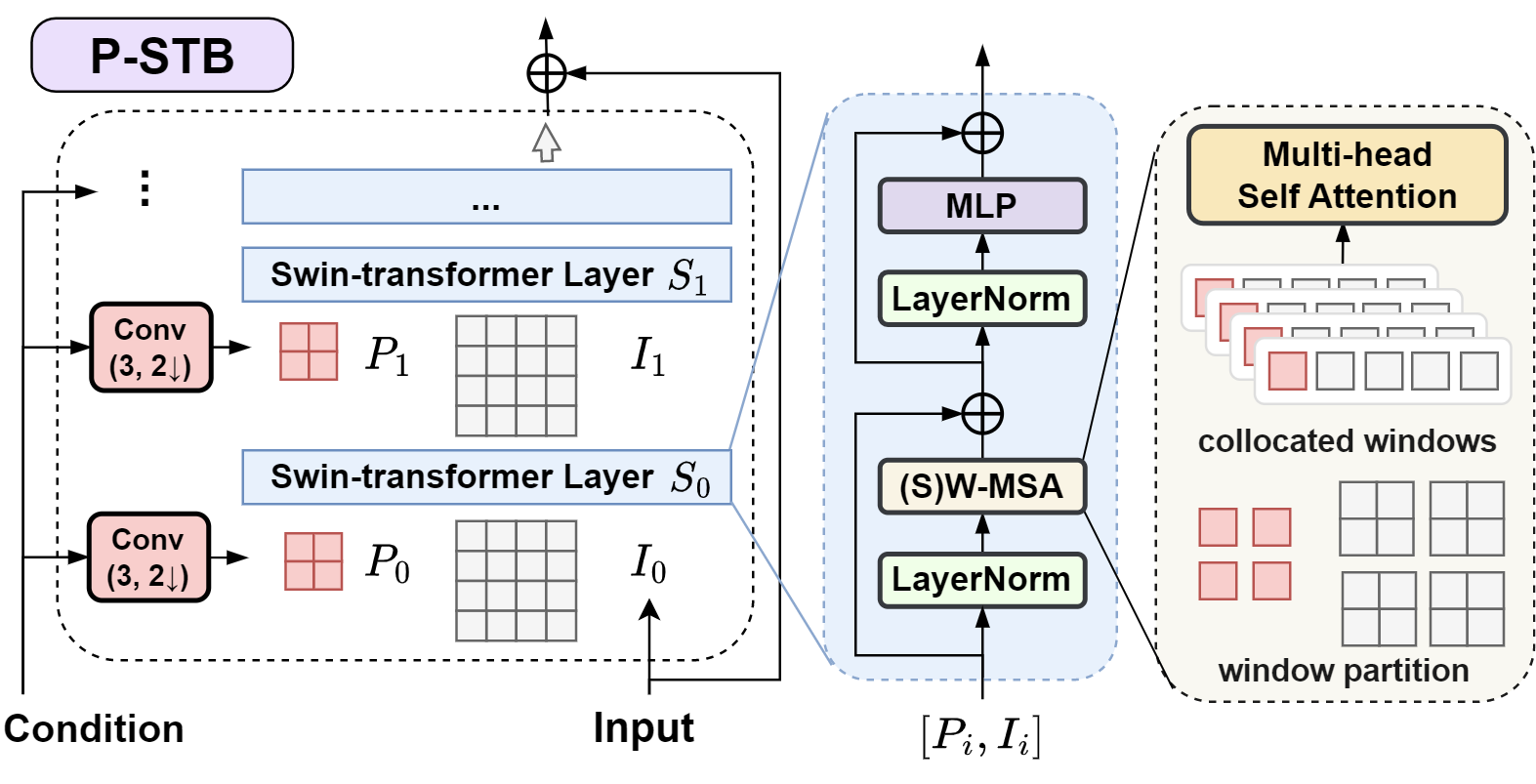}
    \caption{Illustration of the prompted Swin-transformer block.}
    \label{fig:pstb}
    \vspace{-0.3cm}
\end{figure}

Fig.~\ref{fig:pstb} further details P-STB, where $P_i,I_i$ denote the prompt and image tokens, respectively. They are fed into the $i^{\text{th}}$ Swin-transformer layer $S_i$ for window-based attention to arrive at $I_{i+1}$. Specifically, each window has its own image and prompt tokens. We divide spatially the prompt tokens in the same way as the image tokens. For multi-head self-attention, the key $K$ and value $V$ matrices, initially composed of only image tokens $\mathbf{X}_I\in\mathbb{R}^{S_I\times d}$, are augmented with the prompt tokens $\mathbf{X}_P\in\mathbb{R}^{S_P\times d}$, where $S_I, S_P$ are the numbers of image and prompt tokens in a window, respectively, and $d$ is the dimension of each token. In symbols, we have 
\begin{equation}
    \label{eq:qkv}
    \begin{aligned}
         Q &= \mathbf{X}_I\mathbf{W}_Q, \\
         K &= [\mathbf{X}_I, \mathbf{X}_P]\mathbf{W}_K, \\
         V &= [\mathbf{X}_I, \mathbf{X}_P]\mathbf{W}_V, \\
    \end{aligned}
\end{equation}
where $[\cdot]$ indicates concatenation along the token dimension, $\mathbf{W}_Q,\mathbf{W}_K,\mathbf{W}_V\in\mathbb{R}^{d\times d}$ project their respective input matrices into query $Q\in\mathbb{R}^{S_I\times d}$, key $K\in\mathbb{R}^{(S_I+S_P)\times d}$, and value $V\in\mathbb{R}^{(S_I+S_P)\times d}$. Then, we have
\begin{equation}
    \label{eq:msa}
    \begin{aligned}
        \text{Attention}(Q,K,V) &= \text{Softmax}(QK^\top/\sqrt{d}+B)V, \\
    \end{aligned}
\end{equation}
where $B$ denotes the relative position bias. 
Due to the use of an additional strided convolution in P-STB, the number of the prompt tokens is only one fourth of that of the image tokens. This helps reduce the complexity. Our design differs from~\cite{jia2022visual}, where all the non-overlapping windows in a Swin-transformer layer share the same learned prompts. We argue that this is not optimal for spatially adaptive quality control such as ROI coding. 
With our design, an output image token in $I_{i+1}$ aggregates information from the input image $I_{i}$ and prompt $P_{i}$ tokens in the same window.


\begin{figure*}
    \begin{subfigure}{0.244\textwidth}
    \centering
    \includegraphics[width=1\textwidth]{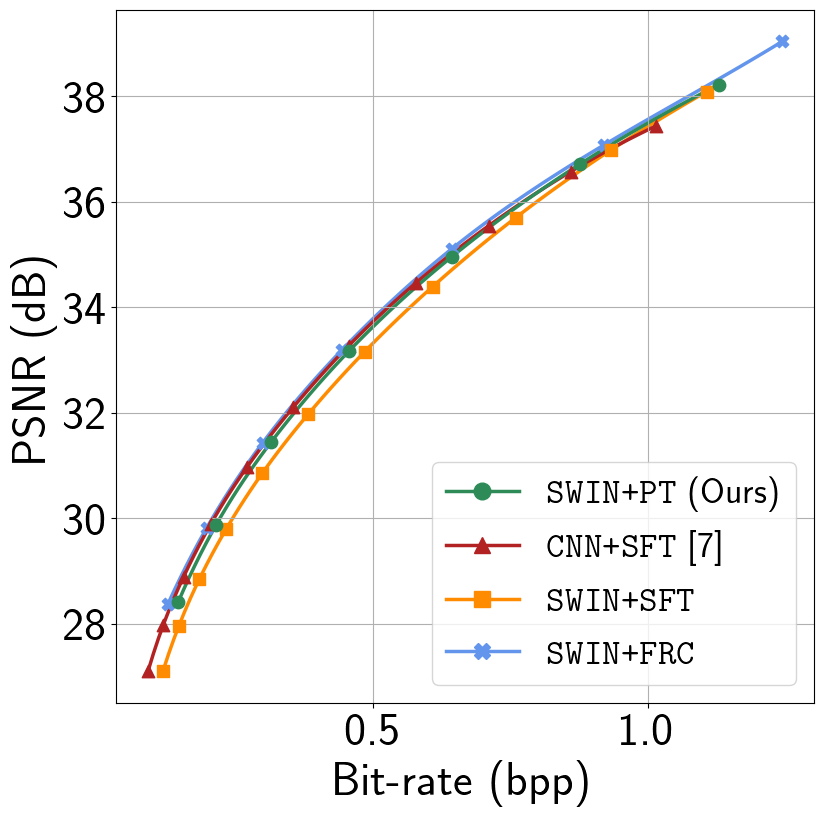}
    \caption{}
    \label{fig:variablerate}
    \end{subfigure}
    \begin{subfigure}{0.244\textwidth}
    \centering
    \includegraphics[width=1\textwidth]{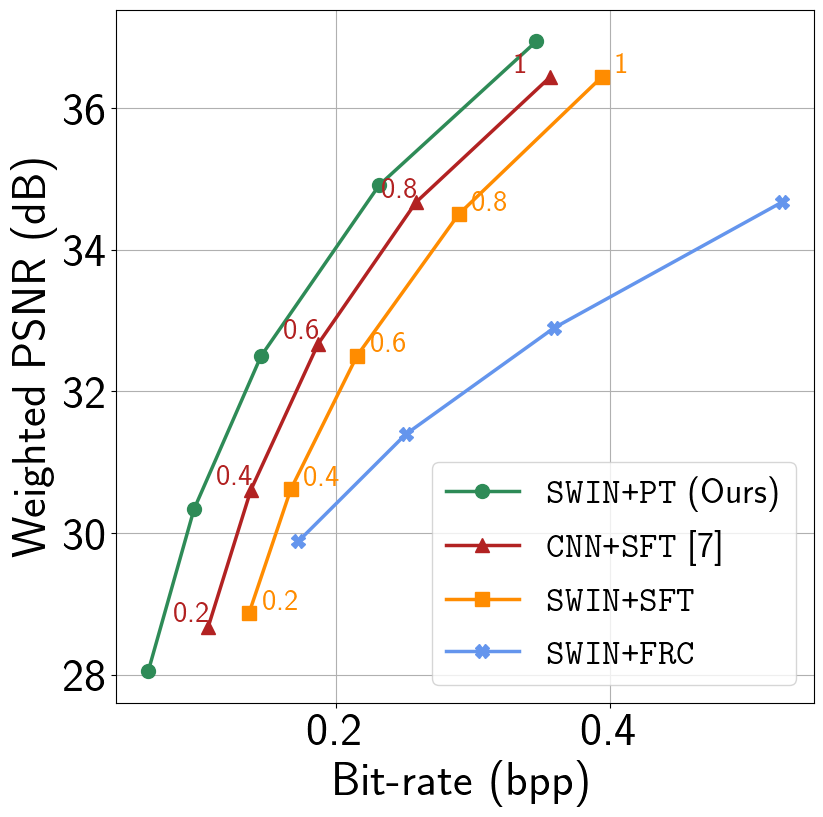}
    \caption{}
    \label{fig:roipsnr}
    \end{subfigure}
    \begin{subfigure}{0.244\textwidth}
    \centering
    \includegraphics[width=1\textwidth]{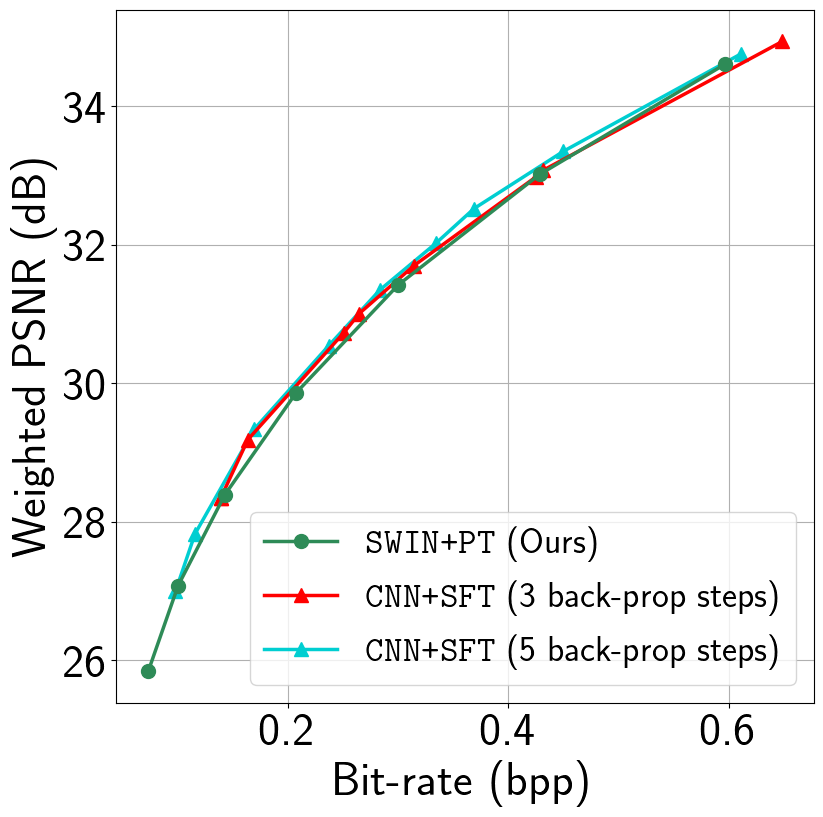}
    \caption{}
    \label{fig:roi_0802}
    \end{subfigure}
    \begin{subfigure}{0.244\textwidth}
    \centering
    \includegraphics[width=1\textwidth]{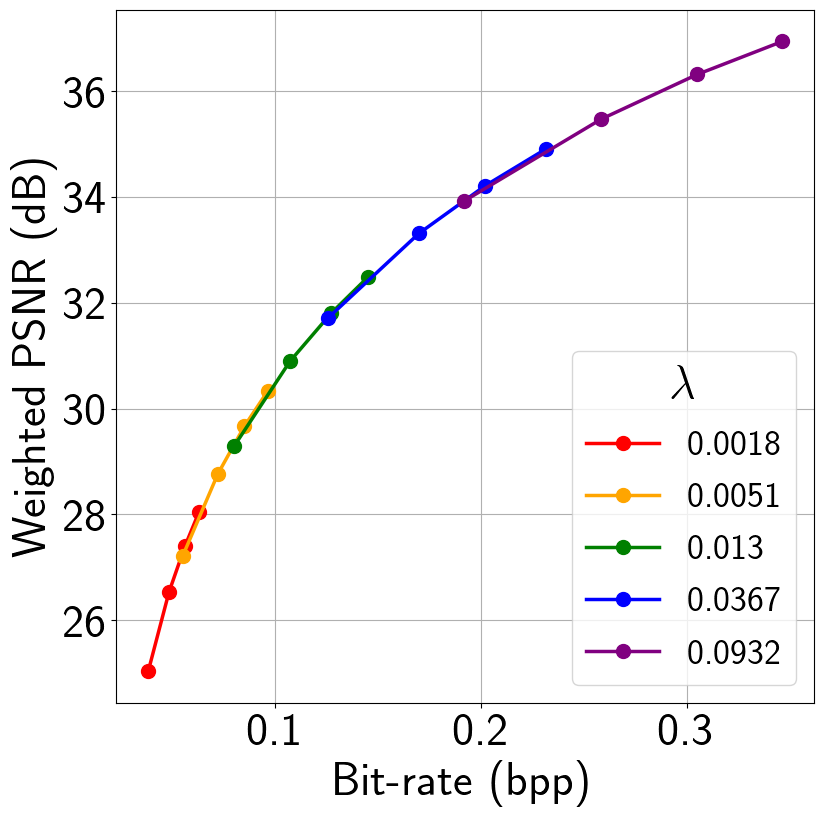}
    \caption{}
    \label{fig:raterange}
    \end{subfigure}
    \vspace{-3mm}
    \caption{(a) Variable-rate coding without ROI on Kodak. (b) Variable-rate coding with ROI on COCO. The annotated numbers indicate the quality values of ROI for $\texttt{CNN+SFT}$ and $\texttt{SWIN+SFT}$. (c) Variable-rate coding optimized for a given ROI specification on 10 randomly selected images from COCO. (d) Rate-distortion plots by altering $\mathbf{M}_\lambda=\lambda$ and $\mathbf{M}_R$ on COCO.}
    \vspace{-0.3cm}
\end{figure*}

\subsection{Loss Function}
We design our loss function in such a way that the model will respond to both the ROI mask $\mathbf{M}_{R_i}$ and the rate parameter $m_{\lambda}$ properly. Specifically, it is given by a weighted sum of the masked distortion and the bit rate: 
\begin{equation}
  \label{eq:roi_rdloss}
  \begin{aligned}
    \mathcal{L}_{rd}(x) = \lambda \cdot \sum\nolimits^{N}_{i=1}\frac{{\mathbf{M}_R}_i \cdot (x_i-x'_i)^2}{N} + \mathcal{R},
  \end{aligned}
\end{equation}
where $x_i,x'_i$ are the $i$-th pixel in the original and compressed images, respectively, $N$ is the total number of pixels in the image, $\mathcal{R}$ denotes the bit rate in bits-per-pixel, and $\lambda = f(m_\lambda) = \exp((\log\lambda_{max}-\log\lambda_{min})\cdot m_\lambda+\log\lambda_{min})$ is a Lagrange multiplier, which is a function of the rate parameter $m_{\lambda}$, with $\lambda_{max}$ and $\lambda_{min}$ being the highest and lowest $\lambda$, respectively. In Eq.~\eqref{eq:roi_rdloss}, the squared error of each pixel is weighted by ROI mask $\mathbf{M}_R$ to reflect its spatial importance, and $\lambda$ trades off the distortion against the bit rate. 
\section{Experiments}
\label{sec:experiment}


\noindent\textbf{Training Details.} We train our model in three stages using Flicker2W~\cite{liu2020unified} and COCO 2017~\cite{lin2014microsoft} training sets. In each training iteration, the input images are randomly cropped to $256\times256$. We first pre-train a base codec (i.e. $g_a, g_s, h_a, h_s$) without the prompt generation networks for the highest rate point. We then train the whole model jointly for variable-rate coding by sampling $\lambda$ uniformly from $\lambda_{min}=0.0018$ to $\lambda_{max}=0.0932$. In this stage, a uniform ROI mask filled with 1's (i.e. every pixel is equally important) is applied. Lastly, we follow~\cite{song2021variable} to produce 4 types of random ROI masks and fine-tune the model for spatial quality control.



\noindent\textbf{Evaluation.} We evaluate our model for variable-rate coding \textit{without} and \textit{with} ROI. In the absence of ROI, we evaluate our model on Kodak~\cite{kodak} by having $\mathbf{M}_R=\mathbf{1}$. In the presence of ROI, we adopt COCO 2017~\cite{lin2014microsoft} validation set for testing. The image reconstruction quality is measured in terms of weighted PSNR, for which the weighted mean-squared error is evaluated by $(\alpha\text{MSE}_\text{ROI} +\beta\text{MSE}_\text{NROI})/(\alpha N_\text{ROI}+\beta N_\text{NROI})$,
where $\alpha, \beta$ are the weighting factors for the ROI and non-ROI regions, respectively, $\text{MSE}_\text{ROI}$ (respectively, $\text{MSE}_\text{NROI}$) is the sum of the squared errors over ROI (respectively, non-ROI), and $N_\text{ROI}$ (respectively, $N_\text{NROI}$) is the number of ROI (respectively, non-ROI) pixels. Under this setting, the ROI is specified by the union of all the foreground objects in the ground-truth segmentation mask.


\noindent\textbf{Baselines.} For comparison, the baseline methods include (1) training separate models for variable-rate coding without $p_a, p_s$ (i.e. separate models for different rates), denoted as \texttt{SWIN+FRC}, and (2) the spatial feature transform (SFT) method in~\cite{song2021variable} (i.e.~\texttt{CNN+SFT}). To validate the effectiveness of prompt tuning, we further construct (3) a model, denoted as \texttt{SWIN+SFT}, that introduces a SFT layer after every STB in our transformer-based codec for variable-rate ROI coding.  


\begin{figure}[t]
    \centering
    \begin{subfigure}{0.43\textwidth}
    \centering
    \includegraphics[width=1\textwidth]{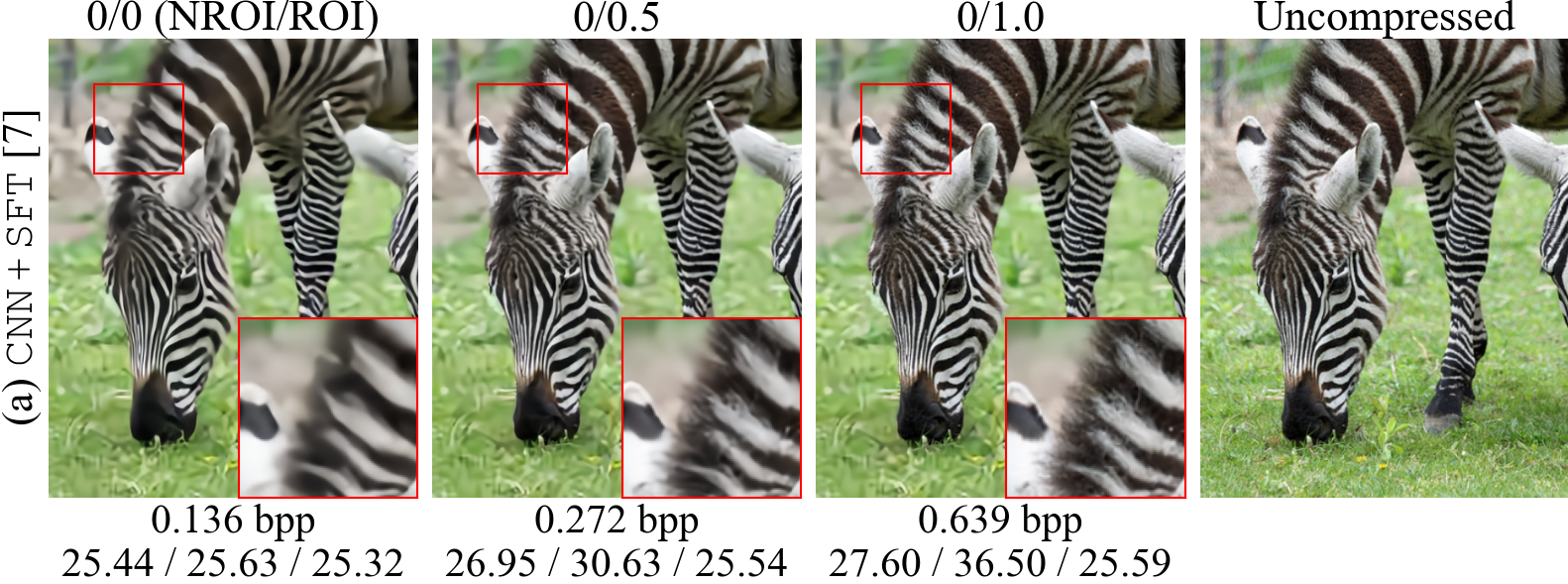}
    \label{fig:vis_ICCV}
    \vspace{-3mm}
    \end{subfigure}
    \begin{subfigure}{0.43\textwidth}
    \centering
    \includegraphics[width=1\textwidth]{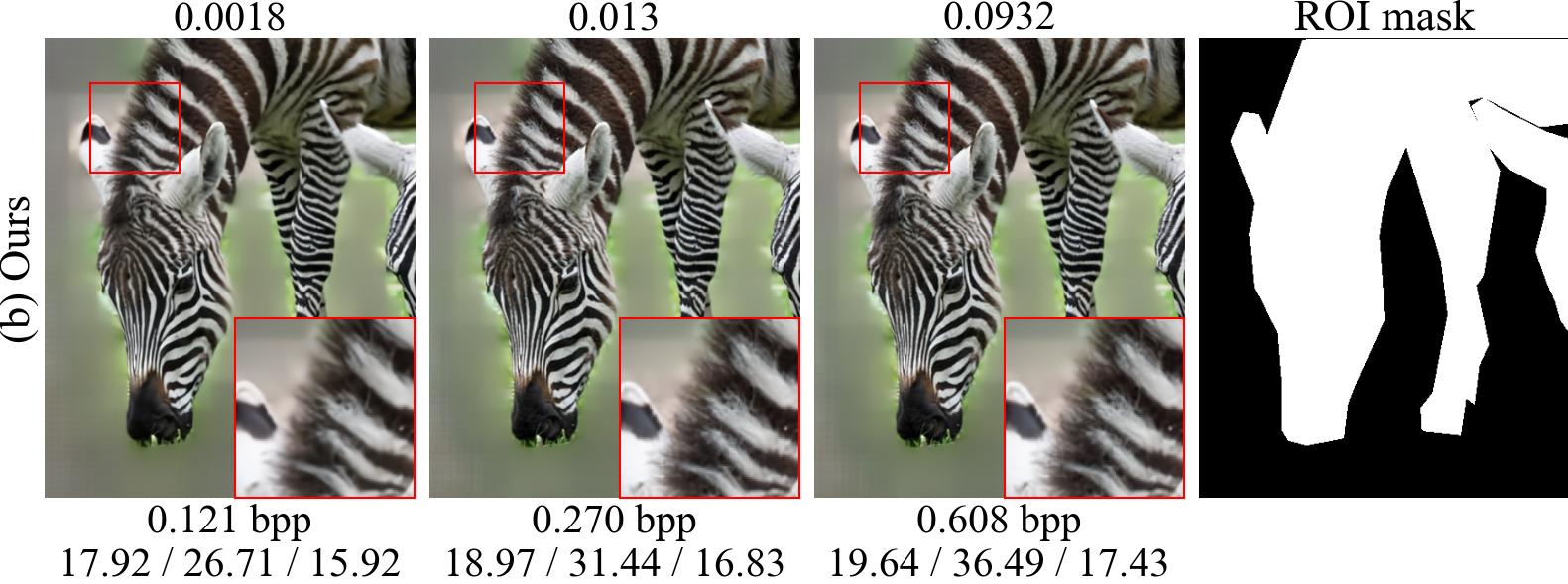}
    \label{fig:vis_ours_fixROI}
    \vspace{-5mm}
    \end{subfigure}
    \caption{Subjective quality comparison of our method and \texttt{CNN+SFT}~\cite{song2021variable}. The corresponding rate and PSNR (full image / ROI / non-ROI) are presented below each image.} 
    \label{fig:vis}
    \vspace{-4mm}
\end{figure}

\subsection{Variable-rate Compression}
\label{sec:exp_variable_rate}
Fig.~\ref{fig:variablerate} compares the competing methods for variable-rate compression without ROI. We see that our method with prompt tuning (\texttt{SWIN+PT}) performs very close to the baseline method with training separate models (\texttt{SWIN+FRC}). In contrast, the variant \texttt{SWIN+SFT} incurs a slight rate-distortion loss. These results suggest that prompt tuning is a more effective approach to variable-rate compression than SFT for our transform-based codec. We also see that \texttt{SWIN+PT} performs comparably to \texttt{CNN+SFT} while having lower computational complexity (Section~\ref{sec:exp_analysis}). 



Fig.~\ref{fig:roipsnr} shows how these methods perform in terms of variable-rate coding with ROI. For this experiment, we set $\alpha=1, \beta=0$ in evaluating the weighted PSNR; that is, we focus only on the quality of the ROI region. Recall that our scheme (\texttt{SWIN+PT}) separates the rate control $M_\lambda$ from the spatial quality control $M_R$. For the present task, $M_R$ is chosen to be the binary ground-truth ROI mask while several distinct $M_\lambda$ values are used for variable-rate compression. Without the disentanglement of the rate and spatial quality control, both \texttt{CNN+SFT} and \texttt{SWIN+SFT} have to rely on adjusting a quality map. Specifically, we fix the quality value of the non-ROI region at 0 and adjust that of the ROI region for rate control. In Fig.~\ref{fig:roipsnr}, our scheme (\texttt{SWIN+PT}) consistently shows higher weighted PSNR than the other baselines. Fig.~\ref{fig:vis} further demonstrates that as compared to \texttt{CNN+SFT}~\cite{song2021variable}, our method (\texttt{SWIN+PT}) is more effective in blurring the background for better foreground coding.

Taking one step further, Fig.~\ref{fig:roi_0802} compares our \texttt{SWIN+PT} with \texttt{CNN+SFT}~\cite{song2021variable} under a more general setting, where the ROI and non-ROI quality is weighted by 0.8 and 0.2, respectively. For \texttt{SWIN+PT}, we simply set $\mathbf{M}_R$ to be 0.8 and 0.2 for the ROI and non-ROI regions, respectively, while adjusting $\mathbf{M}_\lambda$ for variable-rate compression. To achieve the same effect with \texttt{CNN+SFT}, we back-propagate an input-specific rate-distortion loss, $\lambda \times (0.8\text{MSE}_\text{ROI} +0.2\text{MSE}_\text{NROI})+R$, to the input quality map because there is no straightforward way to determine its values for simultaneous rate and spatial quality control. In Fig.~\ref{fig:roi_0802}, our feed-forward-based approach achieves nearly identical rate-distortion performance to the more complicated back-prop-based optimization for \texttt{CNN+SFT} (with 3 and 5 update steps).      

\begin{figure}
    \includegraphics[width=0.48\textwidth]{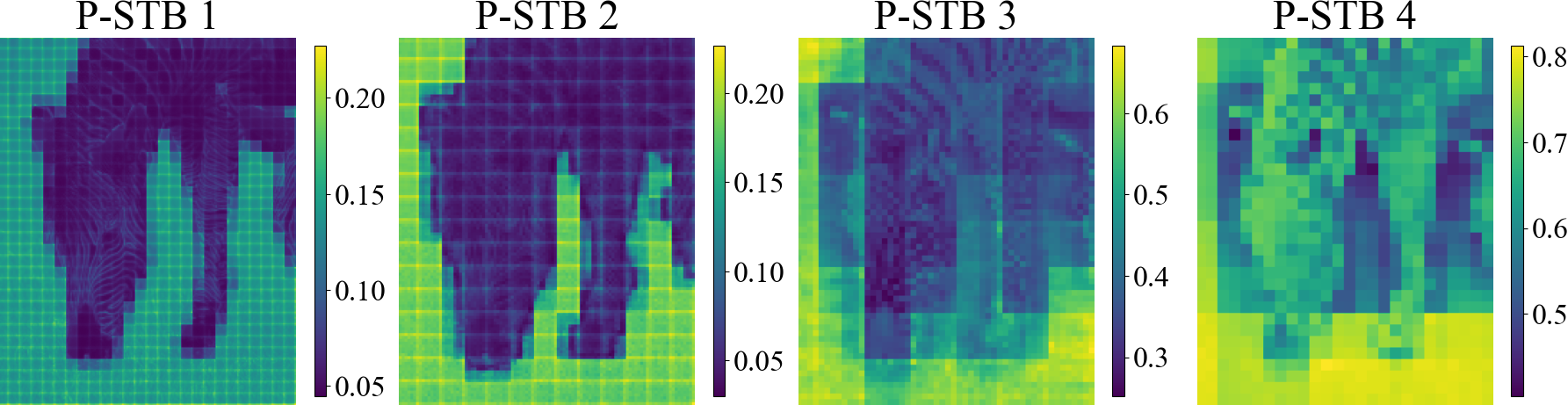}
    \caption{Visualization of the attention map in $S_0$ of each P-STB in $g_a$ with a fixed binary $\mathbf{M}_R$ at $\lambda=0.013$.}
    \label{fig:analysis_var}
    \vspace{-0.1cm}
\end{figure}

\begin{table}
\caption{Comparison of the kMACs/pixel and model size.}
\vspace{-1mm}
\centering\fontsize{9}{9}\selectfont
\begin{tabular}{c||c|c}
\hline
              & kMACs/pixel & Params (M) \\ 
\hline
\texttt{SWIN+FRC}           & 718.50     & 17.66  \\ 
\texttt{CNN+SFT}~\cite{song2021variable}            & 1480.16    & 27.56  \\ 
\texttt{SWIN+SFT}           & 1915.80    & 21.71  \\ 
Ours              & 1070.68  & 32.7   \\ 
\hline
\end{tabular}
\label{tab:complexity}
\vspace{-0.2cm}
\end{table}

\subsection{Analyses of Prompt-based Conditioning}
\label{sec:exp_analysis}

\noindent\textbf{Rate and Spatial Quality Disentanglement.} Fig.~\ref{fig:raterange} demonstrates the effectiveness of our disentanglement of the rate and spatial quality control. The rate-distortion segments of different colors correspond to different choices of $\mathbf{M}_\lambda$. The rate-distortion points of the same colored segment are obtained by setting $\mathbf{M}_R$ in the ROI region to 0.25, 0.5, 0.75, and 1 (and to 0 otherwise). We see that $\mathbf{M}_\lambda$ determines where the major rate point is while $\mathbf{M}_R$ contributes to local bit-rate variations. 

\noindent\textbf{Attention Maps.} Fig.~\ref{fig:analysis_var} visualizes the attention maps for image tokens in different P-STB blocks of the encoder. Each attention map reveals how the prompt tokens attend collectively to every image token. That is, for every image token, we visualize the weighting factors summed over all prompt tokens in the same window. We see that the prompt tokens contribute more distinctively to the ROI and non-ROI regions in the P-STB blocks closer to the input image. In the deeper layers, the distinction between the ROI and non-ROI regions becomes less obvious. 

\noindent\textbf{Complexity Comparison.} Table~\ref{tab:complexity} compares the multiply-accumulate-operations per pixel (kMACs/pixel) and model size of different methods. Even though our model is larger than \texttt{CNN+SFT}~\cite{song2021variable} and \texttt{SWIN+SFT} due to the higher number of channels in $p_a, p_s$, it has lower kMACs/pixel. This is because we process the conditioning signal in lower-resolution feature maps due to a smaller number of prompt tokens, while~\cite{song2021variable} needs to generate affine parameters in the same resolution as the input image. 
Note that our variable-rate model can be more cost-effective than~\texttt{SWIN+FRC}, for which separate models must be trained for different bit rates. That is, its effective model size is a multiple of 17.66M depending on the number of supported rate points.


\vspace{-1mm}
\section{Conclusion}
\label{sec:conclusion}
This work proposes a transformer-based image compression system. It features prompt generation networks to adapt the autoencoder for simultaneous variable-rate and ROI coding. The major finding is that our content-adaptive prompt tuning is more effective than spatial feature transform (SFT) in terms of adapting the transformer-based autoencoder. It also incurs lower kMACs/pixel than SFT.   





\vfill\pagebreak

\bibliographystyle{IEEEbib}
\bibliography{strings,refs}

\end{document}